\documentclass[12pt,letterpaper,nofootinbib,tightenlines]{revtex4}
\usepackage{amssymb,amsmath}
\usepackage{psfrag,verbatim}
\usepackage[dvips]{graphicx}
\usepackage{color}
\definecolor{darkblue}{rgb}{0,0,0.5}
\usepackage[dvips, unicode, colorlinks, bookmarks=true, citecolor=darkblue, linkcolor=darkblue, urlcolor=darkblue]{hyperref}

\newcommand{\hence}{\ \Rightarrow\ }
\newcommand{\spinor}[2]{\begin{pmatrix}#1\\#2\end{pmatrix}}

\newcommand{\spmatrix}[4]{\begin{pmatrix}#1 & #2\\#3 & #4\end{pmatrix}}

\begin{document}

\begin{abstract}

Sensitivity of future laser interferometric gravitational-wave detectors can be improved using squeezed light with frequency-dependent squeeze angle and/or amplitude, which can be created using additional so-called filter cavities. Here we compare performances of several variants of this scheme, proposed during last years, assuming the case of a single relatively short (tens of meters) filter cavity suitable for implementation already during the life cycle of the second generation detectors, like Advanced LIGO. Using numerical optimization, we show that the phase filtering scheme proposed by Kimble et al \cite{02a1KiLeMaThVy} looks as the best candidate for this scenario.

\end{abstract}

\title{Optimal configurations of filter cavity in future gravitational-wave detectors}

\author{F.Ya.Khalili}

\email{farid@hbar.phys.msu.ru}

\affiliation{Physics Faculty, Moscow State University, Moscow 119992, Russia}

\date{Draft of \today}

\maketitle


\section{Introduction}

It is well known that sensitivity of optical interferometric displacement meters can be improved by using squeezed  quantum states of the optical field. In particular, in the case of the Fabry-Perot/Michelson topology, used in contemporary laser interferometric gravitational-wave detectors LIGO \cite{LIGOsite} VIRGO \cite{VIRGOsite}, GEO-600 \cite{GEOsite}, and TAMA \cite{TAMAsite}, squeezed state inside the interferometer can be created by injection of squeezed vacuum into the interferometer dark port \cite{Caves1981}. Depending on the squeeze angle, whether phase or amplitude fluctuations of light can be suppressed. The former tuning reduces the {\em measurement noise}, known also as {\em shot noise}, which spectral density is inversely proportional to the optical power $I_c$ circulating in the interferometer arms. However, it increases the {\em back action}, or {\em radiation-pressure} noise, that is a random force acting on the test mass(es). This noise spectral density is directly proportional to $I_c$. Use of amplitude squeezed vacuum increases the measurement noise and reduces the back action one.

In the contemporary laser interferometric gravitational-wave detectors \cite{Waldman2006, Acernese2006, Hild2006, Ando2005}, the optical power is relatively low and thus of the two quantum noise sources, only measurement noise, that dominates at higher frequencies, affects the detectors sensitivity. The low frequency band is dominated by noise sources of non-quantum origin (most notably by the seismic noise) which are several orders of magnitude larger than quantum back action noise. In this case, the overall sensitivity can be improved by using light with squeezed phase fluctuations. This method is being implemented in GEO-600  currently \cite{Vahlbruch2009} and  will quite probably be implemented in LIGO in a few years \cite{Mavalvala2009}, thanks to the recent achievements in preparation of light squeezed in the working band of contemporary gravitational-wave detectors (10-10000\,Hz) \cite{McKenzie2004, Vahlbruch2006}.

In the planned second generation detectors \cite{Thorne2000, Fritschel2002, Smith2009, AdvLIGOsite} the circulating power will be higher by several orders of magnitude, and technical noises should be reduced significantly. Therefore, the second-generation detectors will be {\it quantum noise limited}: at higher frequencies, the sensitivity will still be limited by shot noise, but at lower frequencies one of the main sensitivity limitation will be radiation-pressure noise. The best sensitivity point, where these two noise sources become equal, is known as the Standard Quantum Limit (SQL) \cite{92BookBrKh}.

In order to obtain sensitivity, better that the SQL, frequency-dependent squeezed light, with phase squeezing at higher frequencies and amplitude squeezing at lower ones, can be used, as was first proposed by Unruh \cite{Unruh1982} and later discussed by several authors in different contexts \cite{87a1eKh, JaekelReynaud1990, Pace1993, 95a1VyZu, 02a1KiLeMaThVy, Willke2002, Harms2003, Buonanno2004, Corbitt2004-3}. The first practical method for generating frequency-dependent squeezed light was proposed by Kimble \textit{et al} \cite{02a1KiLeMaThVy}. They have shown that the necessary dependence can be created by reflecting an ordinary frequency-independent squeezed vacuum (before its injection into the interferometer) from additional properly {\it detuned} filter cavities. This method is known as {\it phase pre-filtering}, because the resulting squeezed state in this case is characterized by the frequency-dependent squeeze angle $\theta(\Omega)$ and the constant squeeze factor $e^{2r}$.

The filter cavities can also be located after the interferometer. In this so-called {\it phase post-filtering} scheme, proposed in \cite{02a1KiLeMaThVy}, the light exiting the interferometer through the dark port is reflected from the filter cavities and then goes to the homodyne detector. This scheme implements, in effect, frequency-dependent homodyne angle. One of the advantages of this scheme is that it does not require squeezing and thus can be used if a squeezed light source is not available.

Yet another scheme, known as {\it amplitude filtering}, was proposed by Corbitt, Mavalvala, and Whitcomb in \cite{Corbitt2004-3}. They suggested to use a {\it resonance-tuned} optical cavity with two partly transparent mirrors as a high-pass filter for the squeezed vacuum. In this scheme, at high frequencies, the phase squeezed vacuum gets reflected by the filter and enters the interferometer such that high-frequency shot noise is reduced; while at low frequencies, ordinary vacuum passes through the filter and enters the interferometer, thus low-frequency radiation-pressure noise remains unchanged. Later it was noted in paper \cite{07a1Kh}, that in this scheme, some information about phase and amplitude fluctuation leak out from the end mirror of the filter cavity, thus degrading the sensitivity. In order to evade this effect, an additional homodyne detection ({\sf AHD}) capturing this information has to be used. This scheme was further developed in \cite{09a1KhMiCh}, where it was proposed to inject additional squeezed vacuum though the filter cavity end mirror and thus suppress also the low-frequencies radiation-pressure noise.

Recently it has been noted that combined amplitude-phase filtering scheme also is possible \cite{Adhikari_private}. In essence, it is the same amplitude filtering scheme with two partly transparent mirrors \cite{Corbitt2004-3}, but with the detuned filter cavity, which creates squeezed light with both squeeze amplitude and angle depending on frequency.

The main technical problem of all these schemes arises due to the requirement that the filter cavities bandwidths should be of the same order of magnitude as the gravitational-wave signal frequency $\Omega\sim10^3\,{\rm s}^{-1}$. The corresponding quality factors have to be as high as $\omega_p/\Omega\sim10^{12}$, where $\omega_p\sim10^{15}\,{\rm s}^{-1}$ is the laser pumping frequency. Therefore, long filter cavities with very high-reflectivity mirrors should be used. In particular, two filter cavities with the same length as the main interferometer arms (4\,Km), placed in the same vacuum chamber side-by-side with the latter ones, was discussed in the article \cite{02a1KiLeMaThVy}. According to estimates made in this paper, the gain in the gravitational wave signals event rate up to two orders of magnitude is feasible is this case, providing $\sim10\,{\rm dB}$ squeezing and/or equivalent increase of the optical power circulating in the interferometer. This design is considered as one of the candidates for implementation in the third generation gravitational wave detectors \cite{LCGTsite, ETsite, 09a1ChDaKhMu}.

On the other hand, it was noted in \cite{Corbitt2004-3, 06a2Kh} that using much less expensive scheme with single relatively short (a few tens of meter, which is comparable with the length of the Advanced LIGO auxiliary mode-cleaner cavities) filter cavity, it is possible to obtain a quite significant sensitivity gain. This scheme does not require any radical changes in the detector design and probably can be implemented during the life cycle of the second generation detectors. The goal of the current paper is to find, which of the several proposed filter cavity options suits best for this scenario.

This paper is organized as follows. In Sec.\,\ref{sec:num}, the schemes to be optimized and the optimization procedure are described. In Sec.\,\ref{sec:results}, the optimization results are presented and discussed.

Appendix \ref{app:noise} contain the explicit equations for the quantum noises of the schemes considered in this paper. These equations are based mostly on the results obtained in the articles \cite{02a1KiLeMaThVy, Corbitt2004-3, 09a1ChDaKhMu} and provided here for the notation consistency and for the reader's convenience. In Appendix \ref{app:lossless}, the particular case of the lossless phase filter cavity is considered, which provides some insight into the relative performance of the two phase filtering schemes.

The main notations and parameter values used in this paper are listed in Table\,\ref{tab:notations}.

\begin{table*}[t]
  \begin{tabular}{|c|c|l|}
    \hline
    Quantity    & Value for estimates                 & Description \\
    \hline
    $\Omega$    &                                     & Gravitational-wave frequency \\
    $c$         & $3\times10^8\,{\rm m/s}$            & Speed of light \\
    $\omega_p$  & $1.77\times10^{15}\,{\rm s}^{-1}$   & Optical pump frequency \\
    $m$         & $40\,{\rm kg}$                      & Test mass \\
    $L$         & $4\,{\rm km}$                       & Interferometer arms length \\
    $\gamma$    &                                     & Interferometer half-bandwidth \\
    $I_c$       & $840\,{\rm kW}$                     & Power circulating in each of the arms \\[1ex]
    $J=\dfrac{8\omega_pI_c}{McL}$ & $(2\pi\times100)^3\,{\rm s}^{-3}$ & \\
    $\eta$      & 0.9                                 & Interferometer effective quantum efficiency \\
    $l$         &                                     & Filter cavity length \\
    $e^r$       & $\sqrt{10}$                         & Input field squeezing factor \\
    $T_I^2$     &                                     & Filter cavity input mirror transmittance\\
    $T_E^2$     &                                     & Filter cavity end mirror transmittance \\
    $A^2$       &                                     & Filter cavity losses per bounce \\
    $\gamma_f$  &                                     & Filter cavity half-bandwidth \\
    $\delta_f$  &                                     & Filter cavity detuning \\
    $\phi$      &                                     & Homodyne angle of main homodyne detector \\
    $\zeta$     &                                     & Homodyne angle of additional homodyne detector \\
    \hline
  \end{tabular}
  \caption{Main notations used in this paper.}\label{tab:notations}
\end{table*}

\section{The schemes and the optimization procedure}\label{sec:num}

\begin{figure*}[t]
  \includegraphics{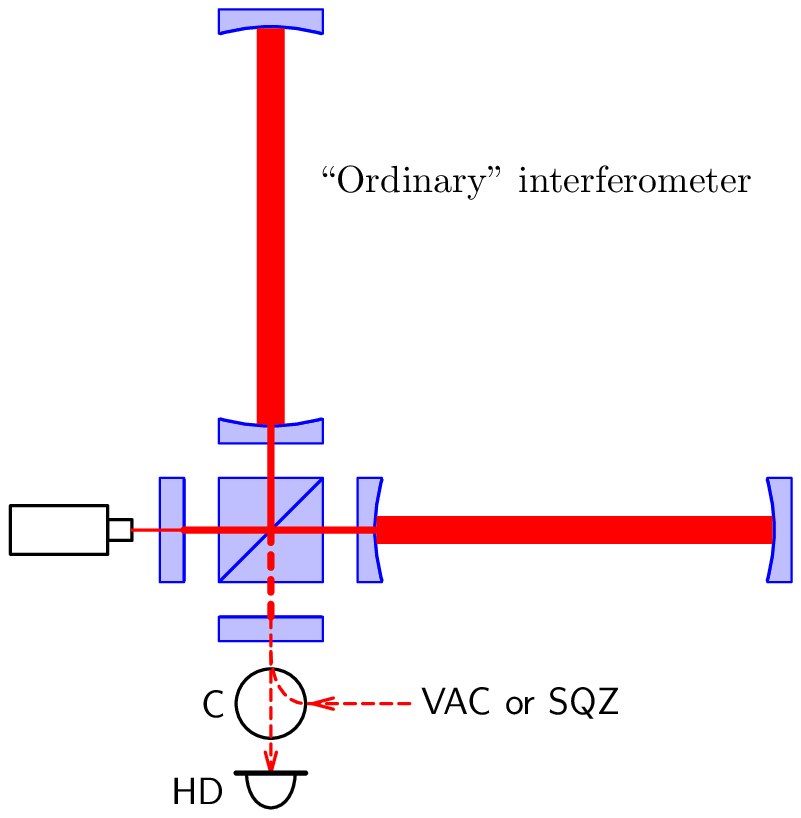}\hfill
  \includegraphics{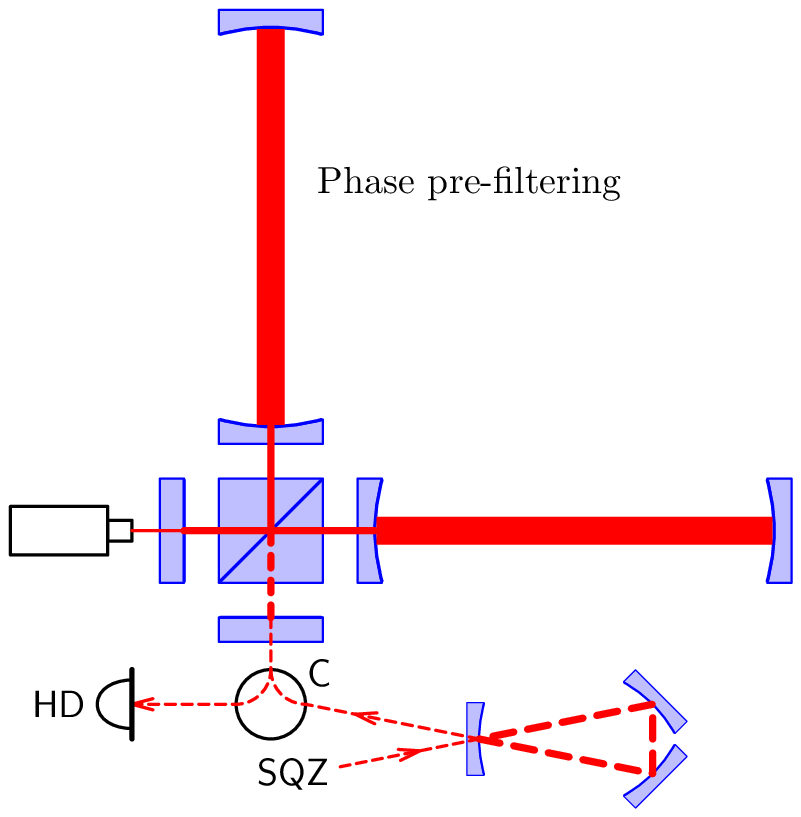}\\[1ex]
  \includegraphics{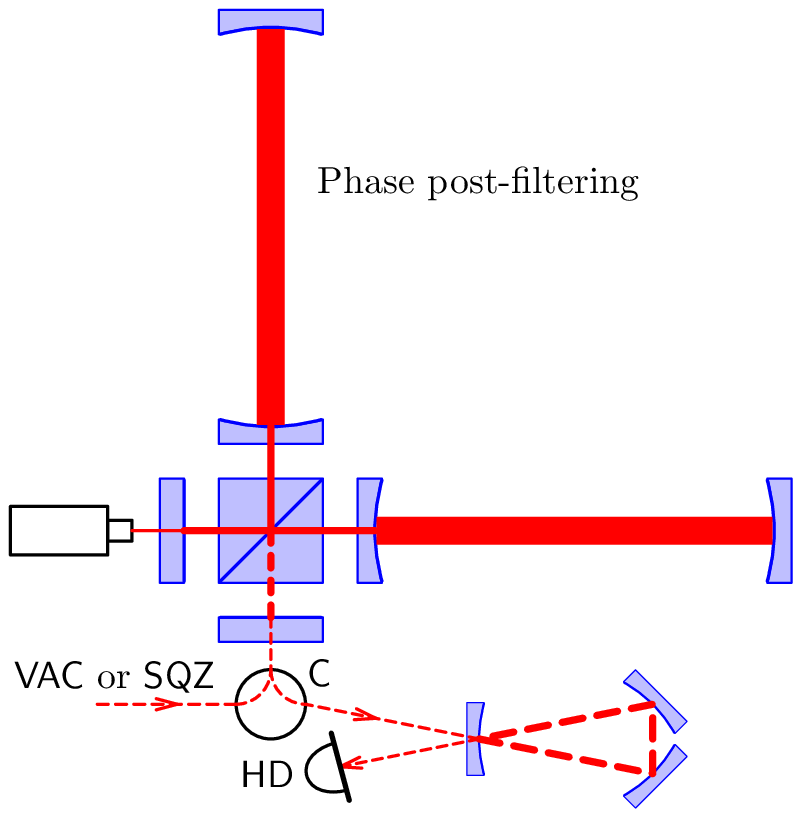}\hfill
  \includegraphics{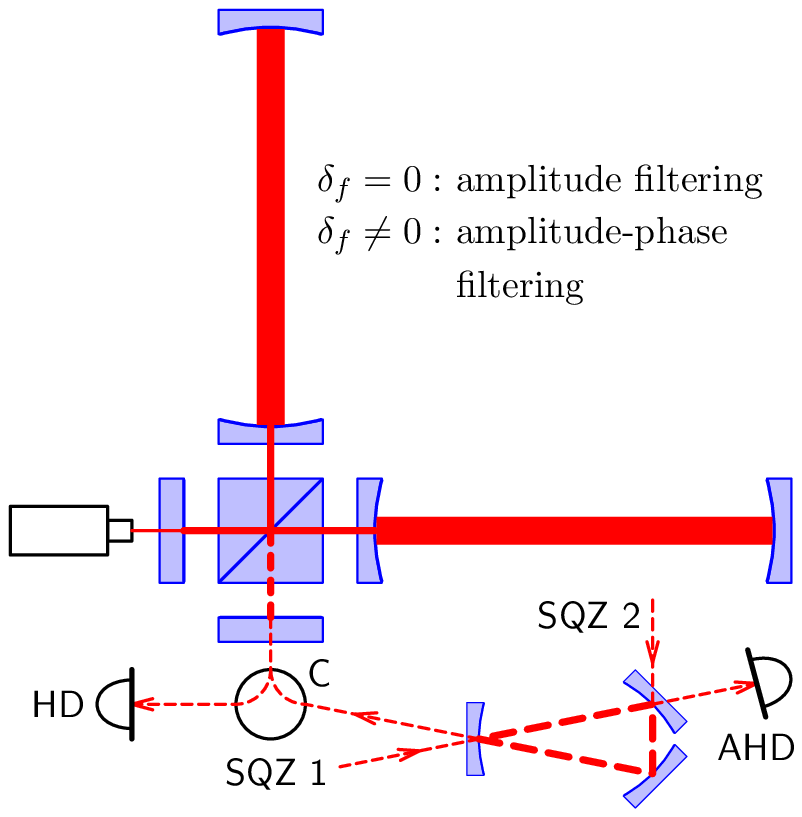}
  \caption{Intereferometers considered in this paper. {\sf HD} --- homodyne detector, {\sf AHD} --- additional homodyne detector, {\sf C} --- some kind of circulator which separate input and output beams ({\it e.g.}, combination of a Faraday rotator and a polarization beam splitter).}\label{fig:topologies}
\end{figure*}

\begin{table*}[t]
  \begin{tabular}{|c|l|c|c|c|c|c|c|c|c|c|}
    \hline 
    & Configuration & $\gamma$ & $\phi$ & $\gamma_I$ & $\gamma_E$ & $\delta_f$ &
       $e^{2r}$ & $\theta_I$ & $\theta_E$ & $\zeta$ \\
    \hline
    1&Ordinary interferometer, vacuum input  & opt & opt & n/a & n/a & n/a & 1 & n/a & n/a & n/a\\
    2&Ordinary interferometer, squeezed input& opt & opt & n/a & n/a & n/a & 10& opt & n/a & n/a\\
    3&Phase post-filtering, vacuum input  & opt & opt & opt & 0   & opt & 1 & $\pi/2-\phi$ & n/a & n/a\\
    4&Phase post-filtering, squeezed input& opt & opt & opt & 0   & opt & 10& $\pi/2-\phi$ & n/a & n/a\\
    5&Phase pre-filtering                 & opt & opt & opt & 0   & opt & 10& $\pi/2-\phi$ & n/a & n/a\\
    6&Amplitude filtering                 & opt & opt & opt & opt & 0   & 10& $\pi/2-\phi$ & opt & opt\\
    7&Combined amplitude-phase filtering  & opt & opt & opt & opt & opt & 10& $\pi/2-\phi$ & opt & opt\\
    \hline
  \end{tabular}
  \caption{Parameters values for the configurations considered in this paper; ``opt'' means that the parameter is optimized in this configuration; ``n/a'' means that the parameter is not applicable to this configuration.}\label{tab:parameters}
\end{table*}

We consider in this paper the following seven configurations, see also Fig.\,\ref{fig:topologies}:

\begin{enumerate}
  \item the ``ordinary'' interferometer (that is, without filter cavity) with vacuum input (no squeezing);
  \item the ``ordinary'' interferometer with squeezed light injection into the dark port;
  \item the phase post-filtering with vacuum input (no squeezing);
  \item the phase post-filtering with squeezed light injection into the dark port;
  \item the phase pre-filtering;
  \item the amplitude filtering;
  \item the combined amplitude-phase filtering.
\end{enumerate}

The first two configurations, which do not contain filter cavity, are included into consideration in order to provide the baseline for the more advanced ones, and to compare the sensitivity gain provided by frequency-independent and frequency-dependent squeezing.

The main interferometer parameters: arms length, mirrors mass, circulating optical power, and optical pump frequency, are assumed to be the same as planned for the Advanced LIGO, see Table \ref{tab:notations}. For the variants, which require squeezed light, we assume 10\,db squeezing. For the filter cavity, we use the following convenient parameters:
\begin{align}\label{gammas}
  & \gamma_I = \dfrac{T_I^2}{4cl} \,, &
  & \gamma_E = \dfrac{T_E^2}{4cl} \,, \\
  & \gamma_L = \dfrac{A^2}{4cl} \,, \label{gamma_L}
\end{align}
which togetether form its half-bandwidth
\begin{equation}
  \gamma_f = \gamma_I + \gamma_E + \gamma_L \,.
\end{equation} 

In Table \ref{tab:parameters}, the parameters used in the optimization procedures for each of the configurations considered in this paper are listed. The number of the optimization parameters varies from 2 for the ordinary interferometer with vacuum input to 7 for the most sophisticated amplitude-phase filtering case. In order to avoid further increase of the parameters space (which is already quite challenging from the computation time point of view), for some of the parameter fixed sub-optimal values, which provide smooth broadband shape of the quantum noise spectral density, are used. Namely, (i) we suppose that the main interferometer is tuned in resonance. In the absence of squeezing and cavities, the interferometer detuning can provide some moderate sensitivity gain \cite{Buonanno2001, Buonanno2002}, but it destructively interferes with other advanced technologies (see, {\it e.g.}, \cite{09a1KhMiCh}). Also, (ii) we suppose the squeeze angle $\theta_I = \pi/2-\phi$; this tuning provides minimum of the shot noise.

We do not consider here technical noises, that is, the mirrors and the suspension thermal noises, seismics, gravity gradient noises {\it etc}, because it is virtually impossible now to predict their level at the later stages of the Advanced LIGO life cycle. It should be noted, however, that the methods considered here provide only relatively modest gain in the quantum noise spectral density, and they do not rely on any deep spectral minima in the quantum noise. Therefore, even equally modest gain in the thermal noise which quite probably will be achieved in the next decade will allow to reach the quantum sensitivity limitations of these schemes.

On the other hand, we take into account optical losses both in the main interferometer and in the filter cavity, as well as the finite quantum efficiency of photodetectors. For the main interferometer and the photodetectors losses, we adopt model of the frequency-independent effective quantum efficiency $\eta$ discussed in Sec.\,2.3 of paper \cite{09a1ChDaKhMu}, and use moderately optimistic value of $\eta=0.95$. \footnote{In signal recycled configurations, losses in the signal recycled cavity, including the beamsplitter absorption, could be the most significant source of frequency dependence of $\eta$. However, estimates show, that for reasonable values of the signal recycling factor, this frequency dependence also can be neglected.}

The filter cavity losses appear in all equations only in combination \eqref{gamma_L} with the filter cavity length $l$. The longer is filter cavity, the less is the influence of losses, for the same value of $A^2$. Therefore, it is convenient to introduce the {\it effective} cavity length as
\begin{equation}
  l_{\rm eff} = \frac{A_0^2}{A^2}\,l \,,  
\end{equation} 
where $A_0^2$ is some fixed value of the losses per bounce. In this paper, we assume, that $A_0^2=10^{-5}$. Therefore, given, for example, a cavity with $l=100\,{\rm m}$ and $A^2=10^{-4}$, the effective length will be equal to $l_{\rm eff}=10\,{\rm m}$.

As the criteria of the optimization, signal-to-noise ratios (SNRs) for the burst sources and for the neutron star-neutron star binary events are used. The first one characterizes broadband sensitivity, while the second is more sensitive to low-frequency noises. It is convenient to normalize the SNR values in terms of those corresponding to some canonical interferometer. Here the ordinary interferometer with vacuum input, homodyne angle $\phi=0$ (so-called classical optimization, which minimizes the shot noise), half-bandwidth $\gamma=J^{1/3}=2\pi\times100\,{\rm s}^{-1}$, and $\eta=1$ (no optical losses) will be used as a canonical one. Thus, the explicit equation for the optimization are the following:
\begin{subequations}\label{SNRS} 
  \begin{gather}
    {\rm SNR}_{\rm burst} = \frac{1}{N_{\rm burst}}
      \int_{f_{\rm min}}^{f_{\rm max}^{\rm burst}}\frac{df}{fS^h(2\pi f)} \,, \\
    {\rm SNR}_{\rm nsns} = \frac{1}{N_{\rm nsns}}
      \int_{f_{\rm min}}^{f_{\rm max}^{\rm nsns}}\frac{df}{f^{7/3}S^h(2\pi f)} \,,
  \end{gather}
\end{subequations}
where $S$ is the quantum noise spectral density to be optimized,
\begin{subequations}
  \begin{gather}
    N_{\rm burst} = \int_{f_{\rm min}}^{f_{\rm max}^{\rm burst}}\!\!\!\frac{df}{fS^h_0(2\pi f)} \,, \\
    N_{\rm nsns} =  \int_{f_{\rm min}}^{f_{\rm max}^{\rm nsns}}\!\!\!\frac{df}{f^{7/3}S^h_0(2\pi f)} \,,
  \end{gather}
\end{subequations}
\begin{equation}
  S^h_0(\Omega) = \frac{h_{\rm SQL}^2(\Omega)}{2}
    \left[\frac{1}{\mathcal{K}(\Omega)} + \mathcal{K}(\Omega)\right]
      \biggr|_{\gamma=2\pi\times100\,{\rm s}^{-1}}
\end{equation} 
is the ``canonical'' interferometer quantum noise spectral density,
\begin{equation}
  h_{\rm SQL}^2(\Omega) = \frac{8\hbar}{ML^2\Omega^2}
\end{equation} 
is the SQL value of the quantum noise spectral density,
\begin{equation}
  \mathcal{K}(\Omega) = \frac{2J\gamma}{\Omega^2(\gamma^2+\Omega^2)} 
\end{equation}
is the optomechanical coupling factor \cite{02a1KiLeMaThVy}, $f_{\rm min}=10\,{\rm Hz}$, $f_{\rm max}^{\rm burst}=10\,{\rm kHz}$ (these two values are defined by the gravitational wave detector bandwidth), and $f_{\rm max}^{\rm nsns}=1.5\,{\rm kHz}$ (See Sec.~3.1.3 of \cite{Postnov2006}. All spectral densities are normalized as equivalent fluctuations of gravitational-wave strain amplitude $h$. The explicit expressions for the optimized spectral densities are provided in the Appendix, see Eqs.\,(\ref{S_vac}, \ref{S_sqz}, \ref{S_post}, \ref{S_pre}, \ref{S_cmwd}).

\section{Discussion}\label{sec:results} 

\begin{figure*}[t]
  \includegraphics[width=0.49\textwidth]{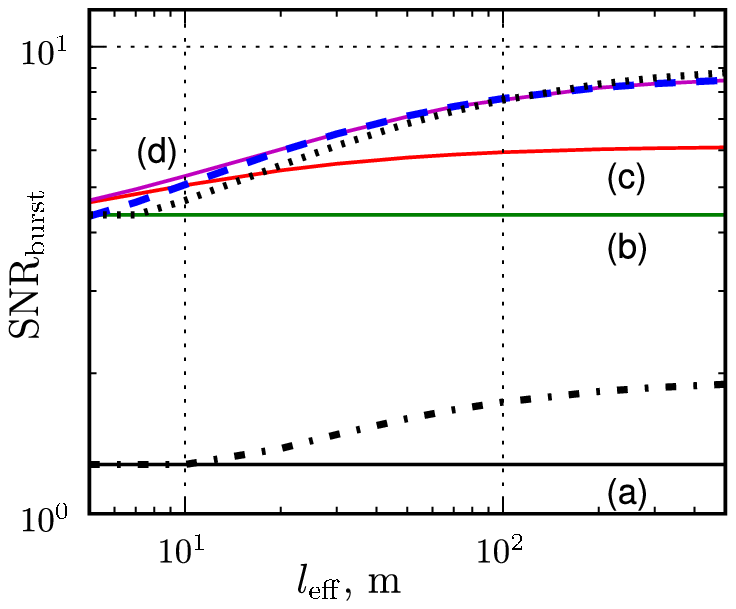}\hfill
  \includegraphics[width=0.49\textwidth]{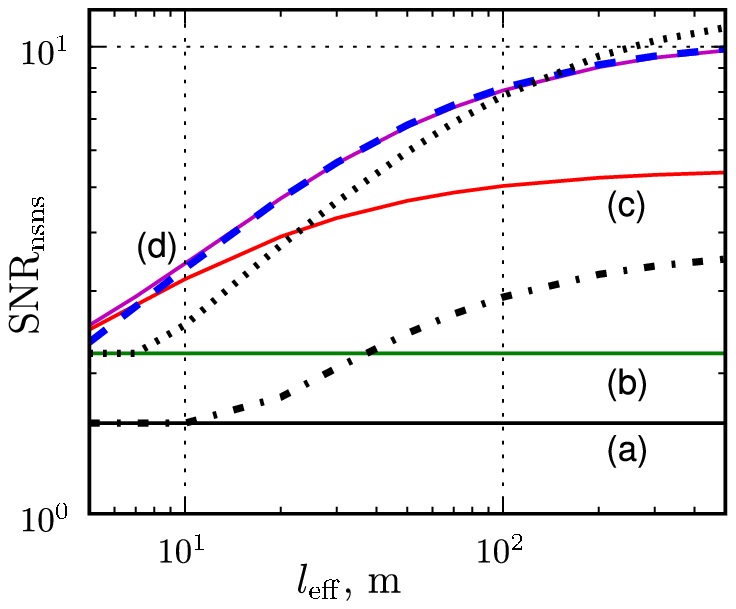}
  \caption{Normalized signal-to-noise ratios as functions of filter cavity effective length. Solid lines: (a) --- ordinary interferometer, vacuum input; (b) --- frequency-independent squeezing; (c) --- amplitude filtering, (d) --- combined amplitude-phase filtering; Dashes: phase pre-filtering; dash-dots: phase post-filtering, vacuum input; dots: phase post-filtering, squeezed input. Left: optimization for bursts; right: optimization for nsns events.}\label{fig:snrs}
\end{figure*}

The results of the numerical optimization are presented in Fig.\,\ref{fig:snrs}, where the SNR values  \eqref{SNRS} are plotted as functions of the effective cavity length $l_{\rm eff}$. The most evident conclusions which follows from these plots are: (i) that sensitivity of the amplitude filtering scheme is inferior to ones of the both phase-filtering variants and (ii) that the results for the combined amplitude-phase filtering scheme are virtually indistinguishable from those for the phase pre-filtering one, except of the very short filter cavity cases, $l_{\rm eff}\lesssim10\,{\rm m}$, where it provides slightly better sensitivity. However, (iii) for such a short filter cavities, the sensitivity is close to one provided by the ordinary frequency-independent squeezing, and the minor additional gain is probably not worth the hassles associated with the filter cavities implementation.\footnote{Due to the additional constrain $\theta_I=\pi/2-\phi$, used in this paper in the filter cavity based schemes optimization (see Table \ref{tab:parameters}), the post-filtering scheme demonstrates even slightly worse sensitivity for $l_{\rm eff}<10{\rm m}$, than frequency-independent squeezing.} On the other hand, (iv) for longer filter cavities, $l_{\rm eff}=50\text{-}500\,{\rm m}$, the sensitivity gain can be very significant, providing the SNR increase (in comparison with frequency-independent squeezing) of $\sim2$ for broadband sources and to $\sim5$ for low-frequency ones, This is equivalent to the event rate increase by a half order of magnitude and almost one order of magnitude, correspondingly.

\begin{figure}[t]
  \includegraphics[width=0.5\textwidth]{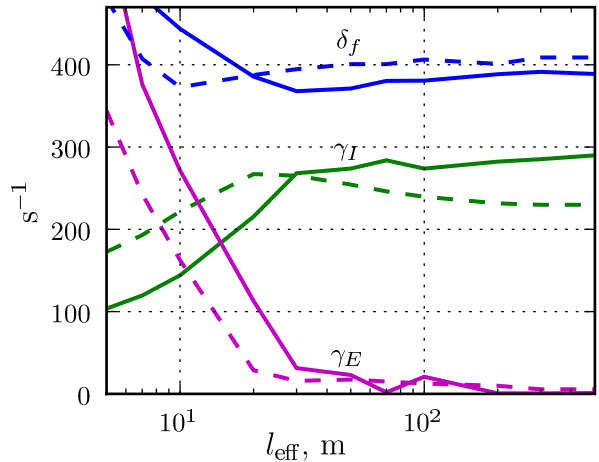}
  \caption{Filter cavity parameters as functions the cavity effective length for the combined amplitude-phase filtering scheme. Solid lines: optimization for bursts, dashed lines: optimization for nsns events.}\label{fig:filter_parms}
\end{figure}

In Fig.\,\ref{fig:filter_parms}, filter cavity parameters for the combined amplitude-phase filtering scheme are plotted as functions of $l_{\rm eff}$. These plots show, why the sensitivity of this scheme is so close to the phase pre-filtering one. If $l_{\rm eff}\gtrsim10\text{-}20\,{\rm m}$, then the optimal transmittance of the filter cavity end mirror quickly drops to zero, while the filter cavity half-bandwidth $\gamma_f\approx\gamma_{fI}$ becomes close to the filter cavity detuning $\delta_f$. These tunings correspond to the phase filtering regime. That is, for the longer cavities, the optimization procedure switches to the pure phase filtering.

\begin{figure*}[t]
  \includegraphics[width=0.49\textwidth]{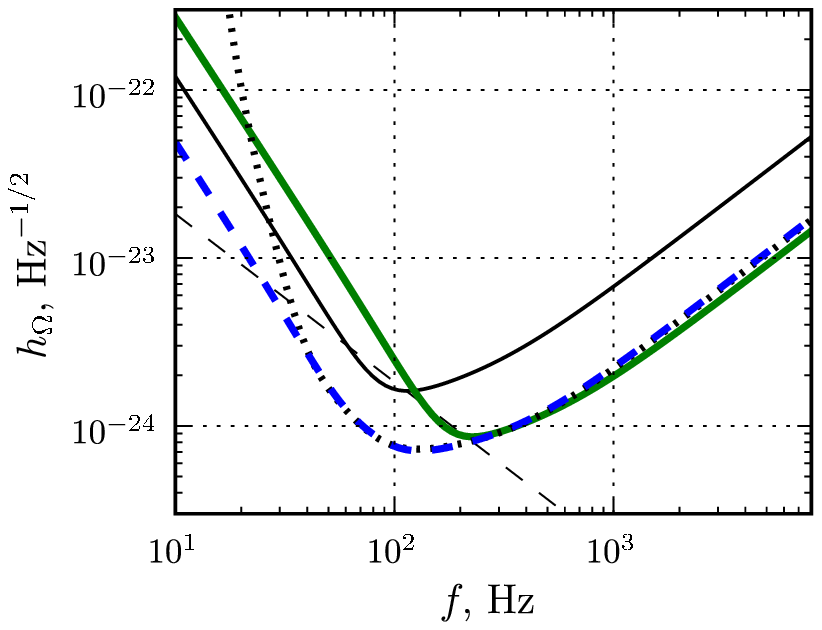}\hfill
  \includegraphics[width=0.49\textwidth]{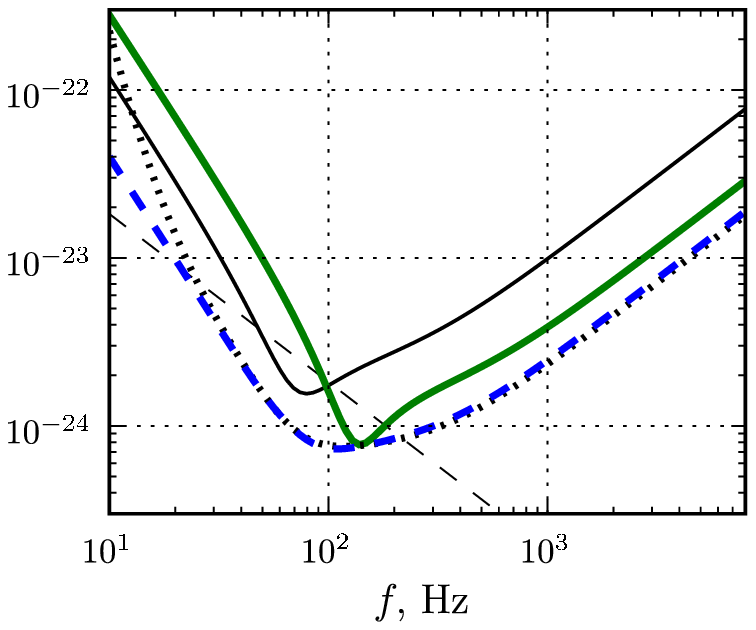}
  \caption{Optimal quantum noises spectral densities. Thin dashes: SQL; thin solid: ordinary interferometer, vacuum input; thick solid: ordinary interferometer, frequency-independent squeezing; thick dashes: phase pre-filtering; dash-dots: phase post-filtering. Left: optimization for bursts; right: optimization for nsns events.}\label{fig:plots}
\end{figure*}

Considering two variants of phase filtering, for the parameters values used here, mostly the pre-filtering scheme demonstrates better results. At a first sight, it looks strange, because it is well known that the post-filtering allows to completely eliminate the back action noise, while the pre-filtering only reduces it by the factor equal to $e^r$ \cite{02a1KiLeMaThVy}. However, the post-filtering scheme is more sensitive to the interferometer losses. It can be explained in the following way. The post-filtering scheme implements frequency-dependent homodyne angle, while the pre-filtering one --- frequency-dependent squeeze angle, compare Eqs.\,(\ref{S_sqz}, \ref{S_post_lossless}, \ref{S_pre_lossless}). In both cases, it allows to measure, at each given frequency, the least noisy quadrature of the output light. However, the homodyne angle affects also the optomechanical transfer factor of the interferometer. At lower frequencies, the post-filtering scheme measures a quadrature which is close to the amplitude one, thus decreasing the transfer factor and emphasizing the additional noise introduced by optical losses. This effect is absent in the pre-filtering scheme, compare the terms proportional to the loss factor $s_{\rm loss}^2$ in Eqs.\,\eqref{S_post_lossless} and \eqref{S_pre_lossless}. As a results, the quantum noise of the post-filtering scheme increases at low frequencies more sharply, than of the pre-filtering one, see Fig.\,\ref{fig:plots}.

Direct comparison of the residual back-action terms (proportional to $\mathcal{K}$) in the optimized quantum noise spectral densities Eqs.\,\eqref{S_post_lossless_opt} and \eqref{S_pre_lossless_opt} for these two scheme allows to conclude, that the post-filtering scheme should be better for the lower losses and not so deep squeezing case, and {\it vice versa}. However, the difference is subtle. More detailed analysis is required here, and the final decision in the post- vs pre-phase filtering choice has to be made with account for additional factors not considered in this paper, in particular, technical noises spectral dependence at low frequencies.

\acknowledgments

This work was supported by NSF and Caltech grant PHY-0651036. The paper has been assigned LIGO document number P0900294.

The author is grateful to Stefan Danilishin and Thomas Corbitt for useful remarks.

\appendix

\section{Quantum noises}\label{app:noise}

\subsection{Some notations}

The following notations are used in this Appendix:
\begin{gather}
  \mathbb{I} = \spmatrix{1}{0}{0}{1} ,  \\
  e^{2i\beta(\Omega)} = \frac{\gamma+i\Omega}{\gamma-i\Omega} \,, \\
  \mathbb{C}(\Omega) =  \spmatrix{1}{\mathcal{K}(\Omega)}{0}{1} , \\
  \Phi(\phi) = \spinor{\cos\phi}{-\sin\phi} , \\
  \mathbb{S}(r,\theta) = \spmatrix{\cosh r + \sinh r\cos2\theta}
    {\sinh r\sin2\theta}{\sinh r\sin2\theta}{\cosh r - \sinh r\cos2\theta} , \\
  \mathbb{Q}(r,\theta) = \mathbb{S}(r,\theta) - \mathbb{I} \,, \\
  s_{\rm loss} = \sqrt{\dfrac{1-\eta}{\eta}} \,.
\end{gather}
Two-photon quadrature amplitude vectors \cite{Caves1985, Schumaker1985} are denoted by boldface letters, and their cosine and sine components --- by the corresponding roman letters with superscripts ``c'' and ``s'', for example:
\begin{equation}
  \hat{\bf a}(\Omega) = \spinor{\hat{\rm a}^c(\Omega)}{\hat{\rm a}^s(\Omega)} \,.
\end{equation}
These components obey the following commutation relations:
\begin{subequations}
  \begin{gather}
    [\hat{\rm a}^c(\Omega), \hat{\rm a}^c(\Omega')] = [\hat{\rm a}^s(\Omega), \hat{\rm a}^s(\Omega')] = 0
    \,, \\
    [\hat{\rm a}^c(\Omega), \hat{\rm a}^s(\Omega')] = 2\pi i\delta(\Omega+\Omega') \,.
  \end{gather}
\end{subequations}
In ground state, they correspond to two independent noises with the one-sided spectral densities equal to 1.

\subsection{``Ordinary'' interferometer}

Using spectral representation and Caves-Schumaker's two-photon formalism \cite{Caves1985, Schumaker1985}, the input-output relations of the noiseless gravitational-wave detector can be presentes as follows \cite{02a1KiLeMaThVy}:
\begin{equation}\label{interf}
  \hat{\bf b}(\Omega)
  = \mathbb{C}(\Omega)e^{2i\beta(\Omega)}\hat{\bf a}(\Omega)
  + \frac{\sqrt{2\mathcal{K}(\Omega)}\,e^{i\beta(\Omega)}}{h_{\rm SQL}(\Omega)}\spinor{h(\Omega)}{0} \,,
\end{equation} 
where $\hat{\bf a}$, $\hat{\bf b}$ are the interferometer input and the output fields and
$h(\Omega)$ is the spectrum of gravitational-wave strain amplitude.

Injection of a squeezed state into the interferometer dark port is described by the following equation:
\begin{equation}\label{vac2sqz} 
  \hat{\bf a}(\Omega) = \mathbb{S}(r,\theta)\hat{\bf z}(\Omega) \,,
\end{equation} 
where operator $\hat{\bf z}(\Omega)$ corresponds to vacuum input field of the squeezer.

Optical losses in the interferometer and the finite quantum efficiency of the photodetector can be taken into account by an imaginary beamsplitter that mixes output of ideal interferometer \eqref{interf} with an additional vacuum noise $\hat{\bf n}$, with weights $\sqrt{\eta}$ and $\sqrt{1-\eta}$:
\begin{equation}\label{loss} 
  \hat{\bf d}(\Omega) = \sqrt{\eta}\,\hat{\bf b}(\Omega) + \sqrt{1-\eta}\,\hat{\bf n}(\Omega) \,.
\end{equation}
The photodetector output signal ({\it i.e.}, the differential current of the homodyne detector) is proportional to
\begin{equation}
  \Phi^+(\phi)\hat{\bf d}(\Omega) \propto h(\Omega) + \hat{h}_{\rm fluct}(\Omega) \,,
\end{equation} 
where
\begin{equation}\label{h_fluct} 
  \hat{h}_{\rm fluct}(\Omega) = \frac{h_{\rm SQL}(\Omega)}{\sqrt{2\mathcal{K}(\Omega)}\,\cos\phi}\,
  \Phi^+(\phi)\Bigl[
    \mathbb{C}(\Omega)\mathbb{S}(r,\theta)\hat{\bf z}(\Omega)e^{i\beta(\Omega)}
    + s_{\rm loss}\hat{\bf n}(\Omega)e^{-i\beta(\Omega)}
  \Bigr]
\end{equation}
is the sum quantum noise with the spectral density equal to
\begin{multline}\label{S_sqz} 
  S^h_{\rm SQZ}(\Omega) = \frac{h_{\rm SQL}^2(\Omega)}{2\mathcal{K}(\Omega)\cos^2\phi}\bigl[
      \Phi^+(\phi)\mathbb{C}(\Omega)\mathbb{S}(2r,\theta)\mathbb{C}^+(\Omega)\Phi(\phi) + s_{\rm loss}^2
    \bigr] \\
  = \frac{h_{\rm SQL}^2(\Omega)}{2}\biggl\{
      \frac{\cosh2r + \sinh2r\cos2(\phi+\theta) + s_{\rm loss}^2}{\mathcal{K}(\Omega)\cos^2\phi} 
       - 2\frac{\cosh2r\sin\phi - \sinh2r\sin(\phi+2\theta)}{\cos\phi} \\
       + \mathcal{K}(\Omega)(\cosh2r - \sinh2r\cos2\theta)
    \biggr\} .
\end{multline}
In the particular case of vacuum input, $r=0$, this spectral density is equal to
\begin{equation}\label{S_vac}
  S^h_{\rm VAC}(\Omega) = \frac{h_{\rm SQL}^2(\Omega)}{2}
    \left[\frac{1}{\eta\mathcal{K}(\Omega)\cos^2\phi} - 2\tan\phi + \mathcal{K}(\Omega)\right] .
\end{equation}

\subsection{Filter cavity}

\begin{figure}
  \includegraphics{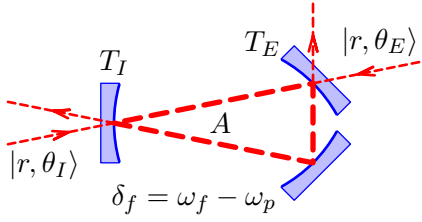}
  \caption{Filter cavity with two input/output ports}\label{fig:filter}
\end{figure}

Input/output relations for the most general case of the detuned filter cavity with two input/output ports shown in Fig.\,\ref{fig:filter} are the following:
\begin{subequations}\label{CMWD_aq}
  \begin{gather}
    \hat{\bf o}_I(\Omega) = \mathbb{R}_I(\Omega)\hat{\bf i}_I(\Omega)
      + \mathbb{T}(\Omega)\hat{\bf i}_E(\Omega) + \mathbb{A}_I(\Omega)\hat{\bf y}(\Omega) \,, \\
    \hat{\bf o}_E(\Omega) = \mathbb{T}(\Omega)\hat{\bf i}_I(\Omega)
      + \mathbb{R}_E(\Omega)\hat{\bf i}_E(\Omega) + \mathbb{A}_E(\Omega)\hat{\bf y}(\Omega) \,,
      \label{CMWD_aq_E} 
  \end{gather}
\end{subequations}
where $\hat{\bf i}_{I,E}$ are the incident fields at the filter cavity input and end mirrors, $\hat{\bf o}_{1,2}$ are the corresponding reflected fileds, $\hat{\bf y}$ is the additional vacuum noise created by absorption in the filter cavity, $\mathbb{R}_{I,E}$ are the reflectivity matrices, $\mathbb{T}$ is the transmittance matrix and $\mathbb{A}_{I,E}$ are the loss matrices. All these matrices have the following uniform structure:
\begin{equation}
  \mathbb{M}(\Omega) = \frac{1}{2}\spmatrix{\mathcal{M}(\Omega) + \mathcal{M}^*(-\Omega)}
    {i[\mathcal{M}(\Omega) - \mathcal{M}^*(-\Omega)]}
    {-i[(\mathcal{M}(\Omega) - \mathcal{M}^*(-\Omega)]}
    {\mathcal{M}(\Omega) + \mathcal{M}^*(-\Omega)} , 
\end{equation} 
where $\mathcal{M} \in \{\mathcal{R}_{I,E},\, \mathcal{T},\, \mathcal{A}_{I,E}\}$ and
\begin{subequations}\label{CMWD_filter} 
  \begin{align}
    {\cal R}_I(\Omega) &= \frac{\gamma_I - \gamma_E - \gamma_L + i(\Omega+\delta_f)}
      {\gamma_f - i(\Omega+\delta_f)} \,, &
    {\cal R}_E(\Omega) &= \frac{\gamma_E - \gamma_I - \gamma_L + i(\Omega+\delta_f)}
      {\gamma_f - i(\Omega+\delta_f)} \,, \\
    {\cal T}(\Omega) &= \frac{-2\sqrt{\gamma_I\gamma_E}}{\gamma_f - i(\Omega+\delta_f)} \,, \\
    {\cal A}_I(\Omega) &= \frac{-2\sqrt{\gamma_I\gamma_L}}{\gamma_f - i(\Omega+\delta_f)} \,,&
    {\cal A}_E(\Omega) &= \frac{2\sqrt{\gamma_E\gamma_L}}{\gamma_f - i(\Omega+\delta_f)}  \,.
  \end{align}
\end{subequations}
Note the following unitarity conditions:
\begin{subequations}\label{CMWD_symm_2}
  \begin{gather}
    \mathbb{R_I}(\Omega)\mathbb{R_I}^+(\Omega) + \mathbb{T}(\Omega)\mathbb{T}^+(\Omega)
      + \mathbb{A}_I(\Omega)\mathbb{A}_I^+(\Omega) = \mathbb{I} \,, \\
    \mathbb{R_E}(\Omega)\mathbb{R_E}^+(\Omega) + \mathbb{T}(\Omega)\mathbb{T}^+(\Omega)
      + \mathbb{A}_E(\Omega)\mathbb{A}_E^+(\Omega) = \mathbb{I} \,, \\
    \mathbb{R}_I(\Omega)\mathbb{T}^+(\Omega) + \mathbb{T}(\Omega)\mathbb{R}_E^+(\Omega)
      + \mathbb{A}_I(\Omega)\mathbb{A}_E^+(\Omega) = 0 \,.
  \end{gather}
\end{subequations}

\subsection{Phase post-filtering}

In the phase filtering cases, both the post-filtering one considered in this subsection and the pre-filtering one considered in the next one, the filter cavity has only one partly transparent mirror:
\begin{equation}
  \gamma_E = 0 \hence \mathbb{T} = 0 \,.
\end{equation} 
In the phase post-filtering case, the interferometer output \eqref{interf} is reflected from the filter cavity, $\hat{\bf i}_I=\hat{\bf b}$, and then goes to the photodetector:
\begin{equation}
  \hat{\bf d}(\Omega) = \hat{\bf o}_I(\Omega)
  = \sqrt{\eta}\bigl[\mathbb{R}_I(\Omega)\hat{\bf b}(\Omega) + \mathbb{A}_I\hat{\bf y}(\Omega)\bigr]
    + \sqrt{1-\eta}\,\hat{\bf n}(\Omega) \,.
\end{equation} 
Therefore, in this case the sum noise is equal to,
\begin{multline}
  \hat{h}_{\rm fluct}(\Omega) = \frac{h_{\rm SQL}(\Omega)}{\sqrt{2\mathcal{K}(\Omega)}}\,
    \left[\Phi^+(\phi)\mathbb{R}_I(\Omega)\spinor{1}{0}\right]^{-1} \\ \times
    \Phi^+(\phi)\Bigl\{
      \mathbb{R}_I(\Omega)\mathbb{C}(\Omega)\mathbb{S}(r,\theta)\hat{\bf z}(\Omega)e^{i\beta(\Omega)}
      + \bigl[
          \mathbb{A}_I(\Omega)\hat{\bf y}(\Omega) + s_{\rm loss}\hat{\bf n}(\Omega)
        \bigr]e^{-i\beta(\Omega)}
    \Bigr\} ,
\end{multline} 
and its spectral density, with account of Eqs.\,\eqref{CMWD_symm_2}, is equal to
\begin{multline}\label{S_post} 
  S^h_{\rm KLMTV\,pre}(\Omega) = \frac{h_{\rm SQL}^2(\Omega)}{2\mathcal{K}(\Omega)}
    \left|\Phi^+(\phi)\mathbb{R}_I(\Omega)\spinor{1}{0}\right|^{-2} \\ \times
    \Bigl\{
      \Phi^+(\phi)\mathbb{R}_I(\Omega)
          \bigl[\mathbb{C}(\Omega)\mathbb{S}(2r,\theta)\mathbb{C}^+(\Omega)-\mathbb{I}\bigr]
        \mathbb{R}_I^+(\Omega)\Phi(\phi)
      + 1 + s_{\rm loss}^2
    \Bigr\} .
\end{multline}

\subsection{Phase pre-filtering}

In the pre-filtering case, squeezed light $\hat{\bf i}_I = \mathbb{S}(r,\theta)\hat{\bf z}$ is reflected from the filter cavity, and then goes to the interferometer dark port:
\begin{equation}
  \hat{\bf a}(\Omega) = \hat{\bf o}_I(\Omega)
  = \mathbb{R}_I(\Omega)\mathbb{S}(r,\theta)\hat{\bf z}(\Omega)
    + \mathbb{A}(\Omega)\hat{\bf y}(\Omega)\,.
\end{equation}
Inserting this $\hat{\bf a}$ into Eq.\,\eqref{h_fluct}, we obtain, that the sum quantum noise in this case is equal to
\begin{multline}
  \hat{h}_{\rm fluct}(\Omega) = \frac{h_{\rm SQL}(\Omega)}{\sqrt{2\mathcal{K}(\Omega)}\,\cos^2\phi}\,
  \Phi^+(\phi)\Bigl\{
    \mathbb{C}(\Omega)\bigl[
      \mathbb{R}_I(\Omega)\mathbb{S}(r,\theta)\hat{\bf z}(\Omega)
      + \mathbb{A}_I(\Omega)\hat{\bf y}(\Omega)
    \bigr]e^{i\beta(\Omega)} \\
    + s_{\rm loss}\hat{\bf n}(\Omega)e^{-i\beta(\Omega)}
  \Bigr\} ,
\end{multline} 
and its spectral density, with account of Eqs.\,\eqref{CMWD_symm_2}, is equal to
\begin{equation}\label{S_pre} 
  S^h_{\rm KLMTV\,pre}(\Omega)= \frac{h_{\rm SQL}^2(\Omega)}{2\mathcal{K}(\Omega)\cos^2\phi}\Bigl\{
    \Phi^+(\phi)\mathbb{C}(\Omega)\bigl[
      \mathbb{R}_I(\Omega)\mathbb{Q}(2r,\theta)\mathbb{R}_I^+(\Omega) + \mathbb{I}
    \bigr]\mathbb{C}^+(\Omega)\Phi(\phi)
    + s_{\rm loss}^2
  \Bigr\} .
\end{equation} 

\subsection{Combined amplitude-phase pre-filtering}

In this case, two squeezed states with different squeeze angles are injected into the filter cavity through two partly transparent mirrors:
\begin{equation}
  \hat{\bf i}_{I,E} = \mathbb{S}(r,\theta_{I,E})\hat{\bf z}_{I,E}(\Omega) \,,
\end{equation}
where $\hat{\bf z}_{I,E}$ are two independent vacuum fields. The field $\hat{\bf o}_I$ then goes to the interferometer dark port:
\begin{equation}
  \hat{\bf a}(\Omega) = \hat{\bf o}_I(\Omega)
  = \mathbb{R}_I(\Omega)\mathbb{S}(r,\theta_I)\hat{\bf z}_I(\Omega)
    + \mathbb{T}(\Omega)\mathbb{S}(r,\theta_E)\hat{\bf z}_E(\Omega)
    + \mathbb{A}_I(\Omega)\hat{\bf y}(\Omega)\,,
\end{equation}
which gives the following equation for the ``naive'' sum quantum noise of interferometer, that is the one which does not take into account entanglement between two outputs of the filter cavity:
\begin{multline}\label{h_ampl} 
  \hat{h}_{\rm fluct}(\Omega) = \frac{h_{\rm SQL}(\Omega)}{\sqrt{2\mathcal{K}(\Omega)}\,\cos\phi}\,
  \Phi^+(\phi)\Bigl\{
    \mathbb{C}(\Omega)\bigl[
      \mathbb{R}_I(\Omega)\mathbb{S}(r,\theta_I)\hat{\bf z}_I(\Omega)
      + \mathbb{T}(\Omega)\mathbb{S}(r,\theta_E)\hat{\bf z}_E(\Omega) \\
      + \mathbb{A}_I(\Omega)\hat{\bf y}(\Omega)
    \bigr]e^{i\beta(\Omega)}
    + s_{\rm loss}\hat{\bf n}(\Omega)e^{-i\beta(\Omega)}
  \Bigr\} .
\end{multline} 
In order to use this entanglement, the field $\hat{\bf o}_E$ has to be detected by an additional homodyne detector. The output signal of this detector is proportional to
\begin{multline}\label{q_ampl}
  \Phi(\zeta)\bigl[\sqrt{\eta}\,\hat{\bf o}_I(\Omega) + \sqrt{1-\eta}\,\hat{\bf n}_a(\Omega)\bigr]
  \propto \hat{q}(\Omega) \\
  = \Phi(\zeta)\bigl[
      \mathbb{T}(\Omega)\mathbb{S}(r,\theta_I)\hat{\bf z}_I(\Omega)
      + \mathbb{R}_E(\Omega)\mathbb{S}(r,\theta_E)\hat{\bf z}_E(\Omega)
      + \mathbb{A}_E(\Omega)\hat{\bf y}(\omega) + s_{\rm loss}\hat{\bf n}_a(\Omega)
    \bigr]
\end{multline} 
where $\zeta$ is the homodyne angle of the additional detector and $\hat{\bf n}_a$ is the noise associated with this detector quantum efficiency which we assume to be also equal to $\eta$.

The optimal combination of both homodyne detectors outputs give the following residual spectral density:
\begin{equation}\label{S_cmwd} 
  S^h_{\rm sum}(\Omega) = S^h(\Omega) - \frac{|S^h_q(\Omega)|^2}{S_q(\Omega)} \,,
\end{equation} 
where [see Eqs.\,\eqref{CMWD_symm_2}]
\begin{subequations}
  \begin{multline}
    S^h(\Omega) = \frac{h_{\rm SQL}^2(\Omega)}{2\mathcal{K}(\Omega)\cos^2\phi}\Bigl\{
      \Phi^+(\phi)\mathbb{C}(\Omega)\bigl[
        \mathbb{R}_I(\Omega)\mathbb{Q}(2r,\theta_I)\mathbb{R}_I^+(\Omega) \\
        + \mathbb{T}(\Omega)\mathbb{Q}(2r,\theta_E)\mathbb{T}^+(\Omega) 
        + \mathbb{I}
      \bigr]\mathbb{C}^+(\Omega)\Phi(\phi)
      + s_{\rm loss}^2
    \Bigr\} \,,
  \end{multline}
  \begin{equation}
    S_q(\Omega) = \Phi^+(\zeta)\bigl[
        \mathbb{T}(\Omega)\mathbb{Q}(2r,\theta_I)\mathbb{T}^+(\Omega)
        + \mathbb{R}_E(\Omega)\mathbb{Q}(2r,\theta_E)\mathbb{R}_E^+(\Omega)
      \bigr]\Phi(\zeta) 
      + 1 + s_{\rm loss}^2 \,, 
  \end{equation}
  \begin{multline}
    S^h_q = \frac{h_{\rm SQL}(\Omega)}{\sqrt{2\mathcal{K}(\Omega)}\,\cos\phi}\,
      \Phi^+(\phi)\mathbb{C}(\Omega)\bigl[
        \mathbb{R}_I(\Omega)\mathbb{Q}(2r,\theta_I)\mathbb{T}(\Omega) \\
        + \mathbb{T}(\Omega)\mathbb{Q}(2r,\theta_E)\mathbb{R}_E(\Omega) 
      \bigr]\Phi(\zeta)e^{i\beta(\Omega)}
  \end{multline}
\end{subequations}
are spectral densities of the noises \eqref{h_ampl}, \eqref{q_ampl}, and their cross-correlation spectral density.

\section{Lossless phase filter cavity}\label{app:lossless} 

In the ideal lossless phase filtering case $\gamma_E=\gamma_L=0$, the transmittance and the loss  matrices vanish, and the refelectivity matrix corresponds to unitary rotation:
\begin{subequations}\label{phase_lossless}
  \begin{gather}
    \mathbb{T}=\mathbb{A}_I=\mathbb{A}_E=0 \,, \\
    \mathbb{R}_I(\Omega) = \sqrt{\frac{\mathbb{D}_f^*(\Omega)}{\mathbb{D}_f(\Omega)}}
      \spmatrix{\cos2\beta_f(\Omega)}{-\sin2\beta_f(\Omega)}
        {\sin2\beta_f(\Omega)}{\cos2\beta_f(\Omega)} \,,
  \end{gather}
\end{subequations}
where
\begin{gather}
  \mathcal{D}_f(\Omega) = (\gamma_f-i\Omega)^2+\delta_f^2 \,, \\
  \cos2\beta_f(\Omega) = \frac{\gamma_f^2-\delta_f^2+\Omega^2}{|\mathcal{D}_f(\Omega)|} \,, \quad
  \sin2\beta_f(\Omega) = \frac{2\gamma_f\delta_f}{|\mathcal{D}_f(\Omega)|} \,.
\end{gather}

With account of Eqs.\,\eqref{phase_lossless}, spectral density \eqref{S_post} can be presented in the form similar to \eqref{S_sqz}, but with frequency-dependent homodyne angle:
\begin{multline}\label{S_post_lossless}
  S^h_{\rm SQZ}(\Omega) = \frac{h_{\rm SQL}^2(\Omega)}{2}\biggl\{
    \frac{\cosh2r + \sinh2r\cos2[\phi_f(\Omega)+\theta] + s_{\rm loss}^2}
      {\mathcal{K}(\Omega)\cos^2\phi_f(\Omega)} \\
     - 2\frac{\cosh2r\sin\phi_f(\Omega) - \sinh2r\sin[\phi_f(\Omega)+2\theta]}
        {\cos\phi_f(\Omega)}
     + \mathcal{K}(\Omega)[\cosh2r - \sinh2r\cos2\theta]
  \biggr\} ,
\end{multline}
where
\begin{equation}
  \phi_f(\Omega) = \phi + 2\beta_f(\Omega) \,.
\end{equation}
If $\phi=0$ and $\theta=\pi/2$, then
\begin{multline}\label{S_post_lossless_simple}
  S^h_{\rm SQZ}(\Omega) = \frac{h_{\rm SQL}^2(\Omega)}{2}\biggl\{
    \frac{1}{{\mathcal{K}(\Omega)}}
      \left[
        e^{-2r} + e^{2r}\tan^22\beta_f(\Omega) + \frac{s_{\rm loss}^2}{\cos^22\beta_f(\Omega)}
      \right] \\
    - 2e^{2r}\tan2\beta_f(\Omega) + \mathcal{K}(\Omega)e^{2r} \biggr\} .
\end{multline}
This spectral density can be minized by setting
\begin{equation}
  \tan2\beta_f(\Omega) = \frac{\mathcal{K}(\Omega)}{1+s_{\rm loss}^2e^{-2r}} \,.
\end{equation} 
With a single filter cavity, this equation can be fulfilled only asymptotically at $\Omega\to0$, by the following values of the filter cavity parameters:
\begin{equation}
  \gamma_f = \delta_f = \sqrt{\frac{J}{\gamma}\,\frac{1}{1+s_{\rm loss}^2e^{-2r}}} \,.
\end{equation}
In this case,
\begin{equation}
  \tan2\beta_f(\Omega) = \frac{2J}{\gamma\Omega^2}\,\frac{1}{1+s_{\rm loss}^2e^{-2r}}  \,,
\end{equation}
and
\begin{equation}\label{S_post_lossless_opt}
  S^h_{\rm SQZ}(\Omega) = \frac{h_{\rm SQL}^2(\Omega)}{2}\biggl[
    \frac{e^{-2r}+s_{\rm loss}^2}{{\mathcal{K}(\Omega)}}
    + \frac{(\Omega/\gamma)^4e^{2r}+s_{\rm loss}^2}{1+s_{\rm loss}^2e^{-2r}}\,\mathcal{K}(\Omega)
  \biggr] .
\end{equation}

In similar way, spectral density \eqref{S_pre}, with account of Eqs.\,\eqref{phase_lossless},  can be presented in the form similar to \eqref{S_sqz}, but with frequency-dependent squeeze angle:
\begin{multline}\label{S_pre_lossless}
  S^h_{\rm SQZ}(\Omega) = \frac{h_{\rm SQL}^2(\Omega)}{2}\biggl\{
    \frac{\cosh2r + \sinh2r\cos2[\phi+\theta_f(\Omega)] + s_{\rm loss}^2}{\mathcal{K}(\Omega)\cos^2\phi}
     \\
    - 2\frac{\cosh2r\sin\phi - \sinh2r\sin[\phi+2\theta_f(\Omega)]}{\cos\phi}
     + \mathcal{K}(\Omega)[\cosh2r - \sinh2r\cos2\theta_f(\Omega)]
  \biggr\} ,
\end{multline}
where
\begin{equation}
  \theta_f(\Omega) = \theta + 2\beta_f(\Omega) \,.
\end{equation}
If $\phi=0$ and $\theta=\pi/2$, then
\begin{multline}\label{S_pre_lossless_simple}
  S^h_{\rm SQZ}(\Omega) = \frac{h_{\rm SQL}^2(\Omega)}{2}\biggl\{
    \frac{\cosh2r - \sinh2r\cos4\beta_f(\Omega) + s_{\rm loss}^2}{\mathcal{K}(\Omega)} 
    - 2\sinh2r\sin4\beta_f(\Omega) \\
    + \mathcal{K}(\Omega)[\cosh2r + \sinh2r\cos4\beta_f(\Omega)]
  \biggr\} .
\end{multline}
This spectral density can be minimized by setting.
\begin{equation}
  \tan2\beta_f(\Omega) = \mathcal{K}(\Omega)\,.
\end{equation}
With single filter cavity, this equation can be fulfilled only asymptotically at $\Omega\to0$, by the following filter cavity parameters:
\begin{equation}
  \gamma_f = \delta_f = \sqrt{\frac{J}{\gamma}} \,.
\end{equation}
In this case,
\begin{equation}
  \tan2\beta_f(\Omega) = \frac{2J}{\gamma\Omega^2} \,,
\end{equation}
and
\begin{equation}\label{S_pre_lossless_opt}
  S^h_{\rm SQZ}(\Omega) = \frac{h_{\rm SQL}^2(\Omega)}{2}\biggl\{
    \left[\frac{1}{\mathcal{K}(\Omega)}+\mathcal{K}(\Omega)\right]e^{-2r}
    + \frac{s_{\rm loss}^2}{\mathcal{K}(\Omega)}
    + \frac{2\mathcal{K}(\Omega)(\Omega/\gamma)^4\sinh2r}
        {1+\mathcal{K}^2(\Omega)[1+(\Omega/\gamma)^2]^2}
  \biggr\} .
\end{equation}

\end{document}